\documentclass[aps,preprint,unsortedaddress]{revtex4-1}
\usepackage{amsmath}% 
\usepackage{graphicx}%
\usepackage{xcolor}%
\usepackage{dcolumn}%
\usepackage{bm}%
\usepackage{ulem}
\usepackage{subcaption}
\usepackage{setspace}
\usepackage{siunitx}
\usepackage{lineno,hyperref}
\usepackage{verbatim} %You can use long comments from this package.
\DeclareMathAlphabet\mathbfcal{OMS}{cmsy}{b}{n}

\begin{document}

\title{Comparison of long-range corrected kernels and range-separated hybrids for excitons in solids}

\author{Rita Maji}
\address{Dipartimento di Scienze e Metodi dell'Ingegneria, Universit{\`a} di Modena e Reggio Emilia,Via Amendola 2 Padiglione Tamburini , I-42122 Reggio Emilia, Italy}
\author{Elena Degoli}
\address{Dipartimento di Scienze e Metodi dell'Ingegneria, Universit{\`a} di Modena e Reggio Emilia and Centro Interdipartimentale En$\&$Tech, Via Amendola 2 Padiglione Morselli, I-42122 Reggio Emilia, Italy\\
Centro S3, Istituto Nanoscienze-Consiglio Nazionale delle Ricerche (CNR-NANO),Via Campi 213/A, 41125 Modena, Italy\\
Centro Interdipartimentale di Ricerca e per i Servizi nel settore della produzione, stoccaggio ed utilizzo dell'Idrogeno H$2$–MO.RE., Via Universit{\`a} 4, 41121 Modena, Italy } 
%}}
\author{Monica Calatayud}
\address{Laboratoire de Chimie Th\'eorique, Sorbonne Universit\'e and CNRS  F-75005 Paris, France}
\author{Val\'{e}rie V\'{e}niard}
\email{valerie.veniard@polytechnique.fr}
\address{Laboratoire des Solides Irradi\'es CNRS, CEA/DRF/IRAMIS, École Polytechnique, Institut Polytechnique de Paris, F-91128 Palaiseau, France and European Theoretical Spectroscopy Facility (ETSF)
}
\author{Eleonora Luppi}
\email{eleonora.luppi@sorbonne-universite.fr}
\address{Laboratoire de Chimie Th\'eorique, Sorbonne Universit\'e and CNRS  F-75005 Paris, France}

\date{\today}

\begin{abstract}

The most accurate theoretical method to describe excitons is the solution of the Bethe-Salpeter equation in the GW approximation (GW-BSE). However, because of its computation cost time-dependent density functional theory (TDDFT) is becoming the alternative approach to GW-BSE to describe excitons in solids. Nowadays, the most efficient strategy to describe optical spectra of solids in TDDFT is to use long-range corrected exchange-correlation kernels on top of GW or scissor-corrected energies. In recent years, a different strategy based on range-separated hybrid functionals started to be developed in the framework of time-dependent generalised Kohn-Sham density functional theory (TDGKSDFT).
Here, we compare the performance of long-range corrected kernels with range-separated hybrid functionals for the description of excitons in solids. This comparison has the purpose to weight the pros and cons of using range-separated hybrid functionals, giving new perspectives for theoretical developments of these functionals. We illustrate the comparison for the case of Si and LiF, representative of solid state excitons.

\end{abstract}

\maketitle

\section{Introduction}

Excitons play a central role in the optical properties of materials for optoelectronics, photovoltaics and photocatalysis applications. \cite{PhysRevB.95.035125,PhysRevLett.116.066803,acs.jpclett.1c00543,Dong2020,science.abm8511} Excitons are usually described as an electron-hole pair and they are classified as Frenkel excitons (bound) localised at the atomic sites and Mott-Wannier excitons (continuum) delocalised over the atomic unit cells. \cite{onid+02rmp}  

An accurate description of the excitonic effect in the optical properties of materials is still very challenging, and nowadays the most accurate theoretical method to describe excitons is the solution of the Bethe-Salpeter equation (BSE) in the GW approximation (GW-BSE). However, the computational cost of GW-BSE can be very high. \cite{reboCRC2013,onid+02rmp}

Time-dependent density functional theory (TDDFT) is an alternative approach to BSE to describe excitons. TDDFT is mathematically simpler than BSE which makes TDDFT computationally more efficient.  The key quantity of TDDFT is the exchange-correlation kernel $f_{\text{xc}}$ which needs to be approximated. Nowadays, none of the proposed approximations for $f_{\text{xc}}$ reaches the BSE accuracy with the only exception of the Nanoquanta exchange-correlation  kernel which, however, makes TDDFT as expensive as BSE. \cite{PhysRevLett.91.056402,PhysRevB.68.165108,PhysRevLett.91.256402}

The most efficient strategy to describe optical spectra of solids in TDDFT is to use long-range corrected exchange-correlation kernels ($1/q^2$ in the long wavelength limit) on top of GW or scissor-corrected energies. \cite{SottIJQC2005,gaurJCP2019} The first long-range corrected (LRC) kernel was derived by Reining {\it et al.} \cite{rein+02prl} through a comparison with BSE. LRC is an empirical kernel which requires a material-dependent parameter. For a large class of semiconductors, the parameter depends on the inverse dielectric constant in a simple way. \cite{bott+04prb} This kernel demonstrated to correctly describe only continuum excitons. Since then, a number of nonempirical exchange-correlation kernels, corrected for the long-range interactions, have been proposed in literature. \cite{PhysRevLett.114.146402,PhysRevLett.107.186401,PhysRevB.87.205143,PhysRevB.95.205136,PhysRevLett.107.186401,PhysRevB.95.205136,PhysRevLett.127.077401} Different efficient bootstrap kernels have been developed, which describe both continuum and strong excitons in insulators and semiconductors. \cite{PhysRevLett.114.146402,PhysRevLett.107.186401} A kernel based on the jellium-with-gap model (JGM) was proposed \cite{PhysRevB.87.205143} to describe both continuum and strong excitons in different materials. However, despite the success of these kernels to describe optical properties, it has been found that they cannot predict accurate exciton binding energies. \cite{PhysRevB.95.205136} In order to recover both properties simultaneously an empirically scaled bootstrap kernel has been proposed. \cite{PhysRevB.95.205136}

In recent years, a different strategy based on range-separated hybrid functionals started to develop. \cite{PhysRevB.78.121201,PhysRevB.92.081204,PhysRevB.92.035202,PhysRevResearch.2.013091,B812838C,ZapJCP2019,Rebo2013MolPhys} Range-separated hybrid functionals rely on the splitting of the Coulomb electron-electron interaction $w_{\text{ee}}=1/r$ into a long-range ($w_{lr}$) and a short-range ($w_{sr}$) contributions by a tunable parameter $\mu$ which controls the range separation. Starting from this simple idea different schemes exist which simulate the $w_{sr}$ by nonlocal Hartree-Fock (HF) exchange and the $w_{lr}$ by (semi-)local density-functional theory exchange functional or vice versa. \cite{HeyScu-JCP-04b,AleScu16JPCL,B812838C,PhysRevB.78.121201}
This methodology has been developed in the framework of the time-dependent generalised Kohn-Sham density functional theory (TDGKSDFT). An important property of TDGKSDFT is that the exchange-correlation potential and the kernel are fully consistent with the choice of the exchange-correlation energy, being its first and second functional derivative with respect to the density. This consistency is not provided by TDDFT with long-range corrected kernels.

The Heyd-Scuseria-Ernzerhof (HSE) range-separated hybrid functional \cite{PhysRevB.78.121201} was used to reproduce optical spectra of semiconductors and insulators. The spectra improve with respect to semilocal functionals for semiconductors showing a very good agreement with experiments, except for large gap insulators. A Coulomb attenuating method (CAM) range-separation was also proposed \cite{PhysRevB.92.081204} showing excellent agreement with both semiconductors and insulators. CAM belongs to range-separated hybrid (RSH) functionals. The main difficulty of these range-separated approaches is to find a general criterium valid for different types of materials for the range-separation parameter $\mu$ and for those parameters that control the weight of nonlocal HF exchange and DFT exchange. To solve this problem, in Ref.(\cite{PhysRevB.92.081204}) the authors optimally tuned the parameters in order to reproduce physical constraints.

Another promising approach is to screen a fraction of the nonlocal HF exchange with the inverse dielectric constant and not to include semilocal exchange-correlation functional. \cite{PhysRevB.92.035202} In this case, however the calculation is not fully consistent as it is obtained from a scissor-corrected local density approximation (LDA) calculation. The same approach was also proposed combining the full range nonlocal HF exchange with a fraction of local exchange functional and correlation. \cite{PhysRevResearch.2.013091}

The use of range-separated hybrid functionals seems to be very promising and open new perspectives for the calculation of optical spectra of solids. However, the calculation of nonlocal HF exchange is computationally more demanding than the standard TDDFT approach with long-range corrected kernels. Moreover, a general rule valid for any materials concerning the choice of the parameters needed in the calculations with hybrid functionals has not been found yet. 

In this paper, we compare the performance of long-range corrected kernels with range-separated hybrid functionals for the description of excitons in solids. This comparison has the purpose to weight the pros and cons of using range-separated hybrid functionals, giving new perspectives for theoretical developments of these functionals. Concerning long-range corrected kernels we studied the LRC \cite{rein+02prl}, scalar RPA bootstrap  (RPA-BO) \cite{PhysRevLett.114.146402} and JGM. \cite{PhysRevB.87.205143}  Concerning hybrid functionals, we investigated the short-range nonlocal HF exchange with and without semilocal exchange-correlation PBE functional \cite{PhysRevLett.77.3865}. We call this two schemes respectively TDHF$^{sr,\mu;\alpha}$ and TDHF$^{sr,\mu;\alpha}$XC$^{\text{PBE}}$. Moreover, in the discussion we also include the hybrid scheme presented in Refs.~\cite{PhysRevB.92.081204,PhysRevMaterials.3.064603} which has a long-range nonlocal HF exchange component.  We illustrate the comparison for the case of Si and LiF, representative of solid state excitons.

In Section \ref{theory} we compare the kernels of BSE, TDDFT with long-range corrected and TDGKSDFT with range-separated hybrid functionals. Section \ref{compdet} is devoted to computational details, while in Section \ref{resultsdiscussion} we present and discuss the results. Conclusions are in Section \ref{conclusion}.% [{\color{magenta} and Appendix in Section ...}]

\section{Theory}
\label{theory}

The macroscopic dielectric tensor $\epsilon_{\text{M}}(\omega)$ is
\begin{equation}
\varepsilon_{\text{M}}(\omega) = \lim_{{\bf q}\to0} \frac{1}{\varepsilon^{-1}_{{\bf G}_1=0,{\bf G}_2=0,}({\bf q}, \omega)}
\label{epsilonm}
\end{equation}
where $\varepsilon^{-1}_{{\bf G}_1,{\bf G}_2}({\bf q}, \omega)$ is the inverse microscopic dielectric matrix written in terms of the reciprocal-space lattice vectors ${\bf G}_1$ and ${\bf G}_2$ for a given wave-vector ${\bf q}$ and frequency $\omega$. The case ${\bf G}_1={\bf G}_2=0$ indicates the head element of the inverse microscopic dielectric matrix.  Through the calculation of $\varepsilon_2(\omega) = \text{Im}[\varepsilon_{\text{M}}(\omega) ]$ the absorption spectrum is obtained. \cite{onid+02rmp}

There exist two different approaches which can be used in GW-BSE, TDDFT and TDGKSDFT to obtain $\varepsilon_{\text{M}}(\omega)$. One approach is the solution of the Dyson equation and the other approach is the solution of the Casida's equations. Converged spectra are identical within the two approaches. \cite{PhysRevB.95.205136} Within these approaches different kernels are used depending on the level of theory.  

The kernel of the BSE in the GW approximation written in reciprocal space and expressed in terms of the 4-space indices ($v{\mathbf k}$, $c{\mathbf k}$, $v'{\mathbf k}'$ and $c'{\mathbf k}'$) of the transition space is \cite{onid+02rmp,PhysRevB.78.121201}
\begin{equation}
\Xi^{\text{GW-BSE}}_{cv{\mathbf k},c'v'{\mathbf k}'} = w_{cv{\mathbf k},c'v'{\mathbf k}'} - W_{c{\mathbf k}c'{\mathbf k}',v{\mathbf k}v'{\mathbf k}'} (\omega) .
\label{gwbse}
\end{equation}
The first term is the Hartree contribution 
\begin{eqnarray}
w_{cv{\mathbf k},c'v'{\mathbf k}'} = \lim_{{\bf q}\to 0}
\sum_{{\bf G}_1{\bf G}_2}\frac{4 \pi}{\vert {\bf q} + {\bf G}_1 \vert^2}\delta_{{\bf G}_1{\bf G}_2} \langle c{\mathbf k} \vert e^{i ({\bf q} + {\bf G}_1) \cdot{\bf r}_1} \vert v{\mathbf k} \rangle \langle v'{\mathbf k}' \vert e^{-i ({\bf q} + {\bf G}_2) \cdot{\bf r}_2} \vert c'{\mathbf k}' \rangle,
\end{eqnarray}
and the second term is the screened Coulomb interaction 
\begin{eqnarray}
W_{c{\mathbf k}c'{\mathbf k}',v{\mathbf k}v'{\mathbf k}'} (\omega) = 4\pi  \lim_{{\bf q}\to 0} \sum_{{\bf G}_1 {\bf G}_2 }
\frac{\varepsilon^{-1}_{{\bf G}_1{\bf G}_2} ( {\bf q},\omega) }{ \vert {\bf q} + {\bf G}_1 \vert^2 }
\langle c{\mathbf k} \vert e^{i ({\bf q} + {\bf G}_1) \cdot{\bf r}_1 }  \vert c'{\mathbf k}' \rangle
\langle v'{\mathbf k}' \vert e^{-i ({\bf q} + {\bf G}_2)\cdot {\bf r}_2} \vert v{\mathbf k} \rangle. 
\label{W}
\end{eqnarray}
where ${\bf q}$ is a vector in the first Brillouin zone, ${\bf G}_1$ and ${\bf G}_2$ are vectors of the reciprocal lattice.

Most of the GW-BSE calculations neglect the frequency dependence of the inverse dielectric function $\varepsilon^{-1}_{{\bf G}_2{\bf G}_1} ( {\bf q},\omega=0)$ which cause the GW-BSE kernel to be static. 

The screened interaction $W$ has a long-range behaviour $1/q^2$ and is attractive, opposite to the Hartree $w^{\text{ee}}$ interaction which is repulsive. We observe that by neglecting $W$ in the kernel we recover the random-phase approximation (RPA). 

The kernel of the TDDFT is  
\begin{equation}
\Xi^{\text{TDDFT}}_{cv{\mathbf k},c'v'{\mathbf k}'} = w_{cv{\mathbf k},c'v'{\mathbf k}'} + f^{\text{xc}}_{cv{\mathbf k},c'v'{\mathbf k}'},
\end{equation}
where the first term is still the Hartree contribution and the second term is the exchange-correlation kernel in the reciprocal and transition space defined as
\begin{eqnarray}
f_{cv{\mathbf k},c'v'{\mathbf k}'}^{\text{xc}} = 
\lim_{{\bf q}\to 0}\sum_{{\bf G}_1{\bf G}_2}f_{\text{xc},{\bf G}_1{\bf G}_2}({\bf q})  \langle c{\mathbf k} \vert e^{i ({\bf q} + {\bf G}_1) \cdot{\bf r}_1} \vert v{\mathbf k} \rangle \langle v'{\mathbf k}' \vert e^{-i ({\bf q} + {\bf G}_2) \cdot{\bf r}_2} \vert c'{\mathbf k}' \rangle.
\label{fxc}
\end{eqnarray}
This quantity is expected to describe the electron correlations that in the BSE are described by $W$. Therefore a good mathematical approximation for the $f^{\text{xc}}$ should include the $1/q^2$ long-range behaviour and the screening. Note also that, by comparing the matrix elements of $W$ in Eq.(\ref{W}) and $f^{\text{xc}}$ in Eq.(\ref{fxc}), there is an exchange between $vk$ and $c'k'$ between $W$ and $f^{\text{xc}}$. The structure of $f^{\text{xc}}$ is similar to the Hartree contribution. In the case of nonlocal HF exchange we obtain exactly the same structure of the matrix elements of $W$ as it will be shown later.

The correct long-range behaviour and screening are described by the long-range corrected kernels. The first of this type of kernels presented in literature is the LRC kernel \cite{rein+02prl,bott+04prb} defined as
\begin{eqnarray}
f^{\text{LRC}}_{\text{xc},{\bf G}{\bf G}'}({\bf q}) = -\frac{\alpha^{\text{LRC}}}{\vert {\bf q}+{\bf G}' \vert^2}\delta({\bf G}',{\bf G}).
\end{eqnarray}
For materials with a small inverse dielectric constant $\varepsilon^{-1}_0$, the parameter $\alpha^{\text{LRC}}$ can be approximated by $\alpha^{\text{LRC}}=4.651\varepsilon^{-1}_0 - 0.213$. \cite{bott+04prb} This kernel has demonstrated to be able to simulate continuum excitons but not strong excitons \cite{rein+02prl,bott+04prb,PhysRevB.95.205136}.

Another long-range corrected kernel is the RPA-BO \cite{PhysRevLett.117.159702} which is defined as
\begin{eqnarray}
f^{\text{RPA-BO}}_{\text{xc}}({\bf q}) = \frac{\varepsilon^{-1}_{\text{RPA},00}w({\bf q})}{1-1/\varepsilon^{-1}_{\text{RPA,00}}({\bf q},0)}.
\end{eqnarray}
This kernel is scalar $({\bf G}=0$ and ${\bf G}'=0$) and the screening is given by the RPA inverse dielectric constant $\varepsilon^{-1}_{\text{RPA},00}$. The RPA-BO kernel gives good results for both continuum and strong excitons in semiconductors and insulators. \cite{PhysRevLett.117.159702}

The JGM kernel \cite{PhysRevB.87.205143,gaurJCP2019} based on the jellium-with-gap model is another kernel with the correct $1/q^2$ behaviour and is defined as
\begin{eqnarray}
f^{\text{JGM}}_{\text{xc},{\bf G}{\bf G}'}({\bf q}) = -4\pi \frac{B'({\bf G} - {\bf G}')}{\vert {\bf q} + {\bf G}'\vert^2} + 4\pi \frac{H({\bf G} - {\bf G}',{\bf G}')}{\vert {\bf q} + {\bf G}'\vert^2} - \frac{D'({\bf G} - {\bf G}')}{1+1/\vert {\bf q}+{\bf G}'\vert^2}
\end{eqnarray}
where $B'$, $H$ and $D'$ depend on the density and on the electronic gap. The precise definition of the quantity is given in  Ref.\cite{PhysRevB.87.205143}. The JGM kernel gives also good results for both continuum and strong excitons in semiconductors and insulators. \cite{PhysRevB.87.205143,gaurJCP2019}

The TDHF$^{sr,\mu;\alpha}$ kernel is 
\begin{equation}
\Xi^{\text{TDHF$^{sr,\mu;\alpha}$}}_{cv{\mathbf k},c'v'{\mathbf k}'} = w_{cv{\mathbf k},c'v'{\mathbf k}'} - \alpha \,w^{\text{HF$^{sr,\mu}$}}_{c{\mathbf k}c'{\mathbf k}',v{\mathbf k}v'{\mathbf k}'}
\label{tdhf}
\end{equation}
and contains the Hartree term and the short-range nonlocal HF exchange 
\begin{eqnarray}
w^{\text{HF$^{sr,\mu}$}}_{c{\mathbf k}c'{\mathbf k}',v{\mathbf k}v'{\mathbf k}'} = 
\lim_{{\bf q}\to 0}\sum_{{\bf G}_1,{\bf G}_2}\frac{4\pi}{\vert {\bf q} + {\bf G}_1 \vert^2}\delta_{{\bf G}_1,{\bf G}_2} 
\langle c{\mathbf k}v'{\mathbf k}' \vert
e^{i ({\bf q} + {\bf G}_1) \cdot({\bf r}_1 - {\bf r}_2)}\text{erfc}(({\bf r}_1 - {\bf r}_2)\mu) \vert v{\mathbf k}c'{\mathbf k}' \rangle.
\end{eqnarray}
screened by a parameter $\alpha$. In the case $\mu=0$ we obtain that  $w^{\text{HF$^{sr,0}$}}$ is equal to nonlocal HF exchange $w^{\text{HF}}$ which is defined as
\begin{eqnarray}
w^{\text{HF}}_{cv{\mathbf k},c'v'{\mathbf k}'} = \lim_{{\bf q}\to 0}\sum_{{\bf G}_1 {\bf G}_2 }
\frac{4\pi }{ \vert {\bf q} + {\bf G}_1 \vert^2 }\delta_{{\bf G}_1,{\bf G}_2} 
\langle c{\mathbf k} \vert e^{i ({\bf q} + {\bf G}_1) \cdot{\bf r}_1 }  \vert c'{\mathbf k}' \rangle
\langle v'{\mathbf k}' \vert e^{-i ({\bf q} + {\bf G}_2)\cdot {\bf r}_2} \vert v{\mathbf k} \rangle,
\label{HF}
\end{eqnarray}
and which has the same matrix form of the unscreened $W$ of the GW-BSE in Eq.(\ref{W}). In this case the kernel is $\Xi^{\text{TDHF$^{sr,0;\alpha}$}} = w- \alpha \,w^{\text{HF}}$. This kernel has been proposed in Refs. \cite{PhysRevB.92.035202} under the name of screened-exact exchange (SXX). In the case of $\mu\to\infty$ we obtain that $w^{sr,\mu}_{\text{HF}}$ is equal to zero and  $\Xi^{\text{TDHF$^{sr,\mu\to\infty;\alpha}$}} = w$ reduces to RPA. 

The kernel TDHF$^{sr,\mu;\alpha}$XC$^{\text{PBE}}$ is
\begin{equation}
\Xi^{\text{TDHF$^{sr,\mu;\alpha}$XC$^{\text{PBE}}$}}_{cv{\bf k},c'v'{\mathbf k}'} = w_{cv{\mathbf k},c'v'{\mathbf k}'} - \alpha \,w^{\text{HF$^{sr,\mu}$}}_{c{\mathbf k}c'{\mathbf k}',v{\mathbf k}v'{\mathbf k}'} + (1-\alpha)f^{\text{x,PBE}}_{cv{\mathbf k},c'v'{\mathbf k}'} + f^{\text{c,PBE}}_{cv{\mathbf k},c'v'{\mathbf k}'}. 
\label{tdhfpbe}
\end{equation}
The same kernel is proposed in Ref.\cite{PhysRevResearch.2.013091}. 
In the case $\mu=0$ we obtain $\Xi^{\text{TDHF$^{sr,0;\alpha}$XC$^{\text{PBE}}$}} = w - \alpha \,w^{\text{HF}} +  (1-\alpha) f^{\text{x,PBE}} +  f^{\text{c,PBE}}$. In the case of $\mu\to\infty$ we obtain $\Xi^{\text{TDHF$^{sr,\mu\to\infty;\alpha}$XC$^{\text{PBE}}$}} = w + (1-\alpha) f^{\text{x,PBE}} + f^{\text{c,PBE}}$. 

In the discussion, we also show the comparison with the range-separated CAM proposed in Refs. \cite{PhysRevMaterials.3.064603,PhysRevB.92.081204} and which also includes a fraction of nonlocal long-range HF exchange. In this case the CAM kernel is
\begin{eqnarray}
\Xi^{\text{TDCAM}^{sr,\mu;\alpha,\beta}}_{cv{\mathbf k},c'v'{\mathbf k}'}  = w_{cv{\mathbf k},c'v'{\mathbf k}'}  - \alpha \,w^{\text{HF$^{sr,\mu}$}}_{c{\mathbf k}c'{\mathbf k}',v{\mathbf k}v'{\mathbf k}'} - (\alpha + \beta)  w^{\text{HF$^{lr,\mu}$}}_{c{\mathbf k}c'{\mathbf k}',v{\mathbf k}v'{\mathbf k}'} \nonumber\\  
+ (1-\alpha)f^{\text{x,PBE$^{sr,\mu}$}}_{cv{\mathbf k},c'v'{\mathbf k}'}  + (1-\alpha-\beta)f^{\text{x,PBE$^{lr,\mu}$}}_{cv{\mathbf k},c'v'{\mathbf k}'} + f^{\text{c,PBE}}_{cv{\mathbf k},c'v'{\mathbf k}'}. 
\end{eqnarray}

This approach requires an additional parameter $\beta$ calculated as $\alpha+\beta=1/\varepsilon_0$ where $\varepsilon_0$ is the material's dielectric constant. \cite{PhysRevMaterials.3.064603,PhysRevB.92.081204} In the case $\mu=0$ we obtain $\Xi^{\text{TDCAM$^{sr,0;\alpha,\beta}$}} = w - \alpha \,w^{\text{HF}} +  (1-\alpha) f^{\text{x,PBE}} +  f^{\text{c,PBE}}$. In the case of $\mu\to\infty$ we obtain $\Xi^{\text{TDCAM$^{sr,\mu\to\infty;\alpha,\beta}$}} = w - (\alpha + \beta)  w^{\text{HF}} + (1-\alpha-\beta) f^{\text{x,PBE}} + f^{\text{c,PBE}}$.

\section{Computational Details}
\label{compdet}

The TDDFT optical spectra with long-range corrected exchange-correlation kernels have been calculated with DP \cite{DP} and 2light \cite{lupp+10jcp,luppi_ab_2010} codes interfaced with the norm-conserving (NC) pseudopotentials and plane-wave basis set ABINIT code. \cite{abinit1,abinit2} 

The optical spectra in GW-BSE and TDGKSDFT with range-separated hybrid functionals have been calculated with the plane-wave based Vienna Ab initio Simulation Package (VASP) with projector augmented-wave (PAW) pseudopotentials. \cite{Hafner, Kresse} 

In the case of NC pseudopotentials we used an energy cutoff of 10 Ha for Si and 40 Ha for LiF, while for PAW pseudopotentials we used an energy cutoff of 9 Ha for Si and 16 Ha for LiF. 

All the calculations have been performed using the experimental lattice parameter 5.430\AA\, for Si and 4.026\AA\, for LiF. We used the experimental lattice parameter in order to be consistent between the different theoretical methods for the spectra comparison.

The convergence parameters for the optical spectra are reported in Table~\ref{conv}. Note that for TDDFT calculations we used shifted k-points grids, while for GW-BSE and TDGKSDFT we averaged the dielectric function over multiple k-points shifted grids.\cite{PhysRevB.78.121201,Hafner,Kresse} \footnote{In the case of GW-BSE/TDGKSDFT, the construction of the Hamiltonian scales as $N_kN^2_vN^2_cN_G$ where $N_k$ is the number of k-points in the Brillouin zone, $N_v$ is the number of valence bands, $N_c$ is the number of conduction bands and $N_G$ is the number of G-vectors. However, except for very large systems, the main limiting factor usually comes from the diagonalization of the Hamiltonian which scales cubically with the matrix rank $N_{rank}^3$ and $N_{rank}=N_kN_vN_c$. TDDFT requires the diagonalization for each k-point of a matrix of rank $N_{rank}=(N_v+N_c)$ and the evaluation of the response function scales as $N_kN_vN_cN_G^2$.}

A broadening of 0.05 eV for all optical spectra have been used.

\begin{table}[h!]
\begin{center}
\begin{ruledtabular}
\caption{Si and LiF convergence parameters. \label{conv}} 
\begin{tabular}{cccccccccc}
%TDDFT (NC) &  \\ 
%\hline
 & Material &  k-points & empty bands &  G-vectors\\
\hline
TDDFT (NC) &Si  & 30$\times$30$\times$30 (shifted)  & 4 &  89 \\
       & LiF    & 32$\times$32$\times$32 (shifted) & 26 & 89 \\
\hline
TDGKSDFT (PAW) &Si & 8$\times$8$\times$8 (29 shifted) & 16 &  163 \\
& LiF          & 8$\times$8$\times$8 (29 shifted) & 32 & 294 \\
\hline
GW-BSE (PAW) & Si & 8$\times$8$\times$8 (29 shifted) & 12/128 & 150 \\
&LiF            & 8$\times$8$\times$8 (29 shifted) & 12/160 & 270 \\

\end{tabular}
\end{ruledtabular}
\end{center}
\end{table}
%-------------------------------------------------------

\section{Results and Discussion}
\label{resultsdiscussion}

The goal of this work is to compare TDDFT and TDGKSDFT to describe optical spectra of solids. As in our calculations we used both NC and PAW pseudopotentials, we have first analysed the electronic structures of Si and LiF. 

For TDGKSDFT, the electronic structure was calculated with the HSE exchange-correlation functional $E^{\text{HSE}^{\mu;\alpha}}_{\text{xc}} = \alpha\,E^{\text{HF}^{\text{sr},\mu}}_{\text{x}} + (1-\alpha) E^{\text{PBE}^{\text{sr},\mu}}_{\text{x}} + E^{\text{PBE}^{\text{lr},\mu}}_{\text{x}} + E_{\text{c}}^{\text{PBE}}$, where the long-range PBE exchange functional is E$^{\text{PBE}^{\text{lr},\mu}}_{\text{x}}$ = E$_{\text{x}}^{\text{PBE}}$ - E$^{\text{PBE}^{\text{sr},\mu}}_{\text{x}}$. \cite{HeyScu-JCP-04b} Considering $\mu\to\infty$ we have $E^{\text{HSE}^{\mu\to\infty;\alpha}}_{\text{xc}}$ = $E^{\text{PBE}}_{\text{xc}}$ and, instead, considering  $\mu=0$ we have $E^{\text{HSE}^{0;\alpha}}_{\text{xc}}$ = $\alpha\,E^{\text{HF}}_{\text{x}} + (1-\alpha)E^{\text{PBE}}_{\text{x}} + E^{\text{PBE}}_{\text{c}}$. Following the optimally tuned strategy, we chose the parameters $\mu$ and $\alpha$ in order to have a good agreement with the GW gaps. For Si we used as ($\mu$;$\alpha$) : (0.2;0.25), (0.3;0.25), (0.3;0.3) and (0.0;0.125), while in the case of LiF we used (0.0;0.4), (0.0;0.45) and (0.0;0.5).

In Table (\ref{Sigaps}) we report the Si gaps calculated in PBE with NC and PAW, together with GW, HSE$^{0.2;0.25}$, HSE$^{0.3;0.25}$, HSE$^{0.3;0.3}$ and HSE$^{0.0;0.125}$ gaps calculated with PAW pseudopotentials. The HSE$^{0.2;0.25}$ gives the closest agreement with the GW gaps. Increasing the value of $\mu$ keeping the value of $\alpha$ constant, as in HSE$^{0.3;0.25}$, has the effect to lower the values of the gaps. This is due to a smaller percentage of nonlocal HF exchange included in the calculation. Instead, increasing the value of $\alpha$ keeping constant the value of $\mu$, as in HSE$^{0.3;0.3}$, has the effect to increase the gap values. In this case a larger amount of nonlocal HF exchange is considered. The HSE$^{0.0;0.125}$ includes a full-range nonlocal HF exchange and the parameter $\alpha$ acts as a screening. A value of $\alpha=0.125$ gives a good agreement with GW gaps.

In Table (\ref{LiFgaps}) we report LiF gaps calculated in PBE with NC and PAW, together with GW and HSE$^{0.0;0.4}$ gaps calculated with PAW pseudopotentials. LiF is a large gap insulator and it requires the correct long-range behaviour of the nonlocal HF exchange. For this reason the HSE performs well only for $\mu=0$. The value of $\alpha=0.4$ was found by imposing the constraints to recover the GW gaps.
%======================================================================
\begin{table*}[h!]
\begin{center}
\begin{ruledtabular}
\caption{Si gaps (eV). The use of norm-conserving pseudopotentials is indicated with the label NC, otherwise PAW pseudopotentials have been used. \label{Sigaps}} 
\begin{tabular}{cccccccccc}
{\text{Si}}      & PBE$^{\text{NC}}$ & PBE & GW & HSE$^{0.2;0.25}$ & HSE$^{0.3;0.25}$ & HSE$^{0.3;0.3}$ & HSE$^{0.0;0.125}$ & Exp\\ 
\hline
$\Gamma_c$ - $\Gamma_v$               &2.58 & 2.57 & 3.34 & 3.33 & 3.15 & 3.27 & 3.28 & 3.35\footnotemark[1]\\
$X_c$ - $\Gamma_v$                  &0.58 & 0.66 & 1.28 & 1.29 & 1.13 &1.22 & 1.33 & 1.17\footnotemark[2]\\
$L_c$ - $\Gamma_v$               & 1.61 & 1.57  & 2.18 & 2.24 & 2.06 & 2.17 & 2.20 & 2.40\footnotemark[3],  2.06\footnotemark[4]\\
\end{tabular}
\end{ruledtabular}
\end{center}
\footnotetext[1]{Reference \cite{PhysRevB.5.497}.}
\footnotetext[2]{Reference \cite{kittel}.}
\footnotetext[3]{Reference \cite{PhysRevLett.54.142}.}
\footnotetext[4]{Reference \cite{HULTHEN19761341}.}
\end{table*}
%===================================================================
\begin{table*}[h!]
\begin{center}
\begin{ruledtabular}
\caption{LiF gaps (eV). The use of norm-conserving pseudopotentials is indicated with the label NC, otherwise PAW pseudopotentials have been used. \label{LiFgaps}} 
\begin{tabular}{ cccccccc } 
{\text{LiF}}      & PBE$^{\text{NC}}$ & PBE & GW & HSE$^{0.0;0.4}$ & Exp\\ 
\hline
$X_c$ - $\Gamma_v$                  & 9.21 & 9.12& 14.21 & 13.99 & 14.20\footnotemark[1]\\
\hline
$X_c$ - $\Gamma_v$                 & 11.19 & 11.25 & 16.36 & 16.12 &\\
\hline
$L_c$ - $\Gamma_v$                  & 13.46 & 13.36  & 19.07 & 18.53 &\\
\end{tabular}
\end{ruledtabular}
\end{center}
\footnotetext[1]{Reference \cite{PhysRevB.13.5530}.}
\end{table*}
%===================================================================

On top of the electronic structure we calculated the optical spectra starting from the lowest level of theory, i.e. the independent-particle approximation (IPA).

\begin{center}
\begin{figure}
\includegraphics[width=0.97\linewidth]{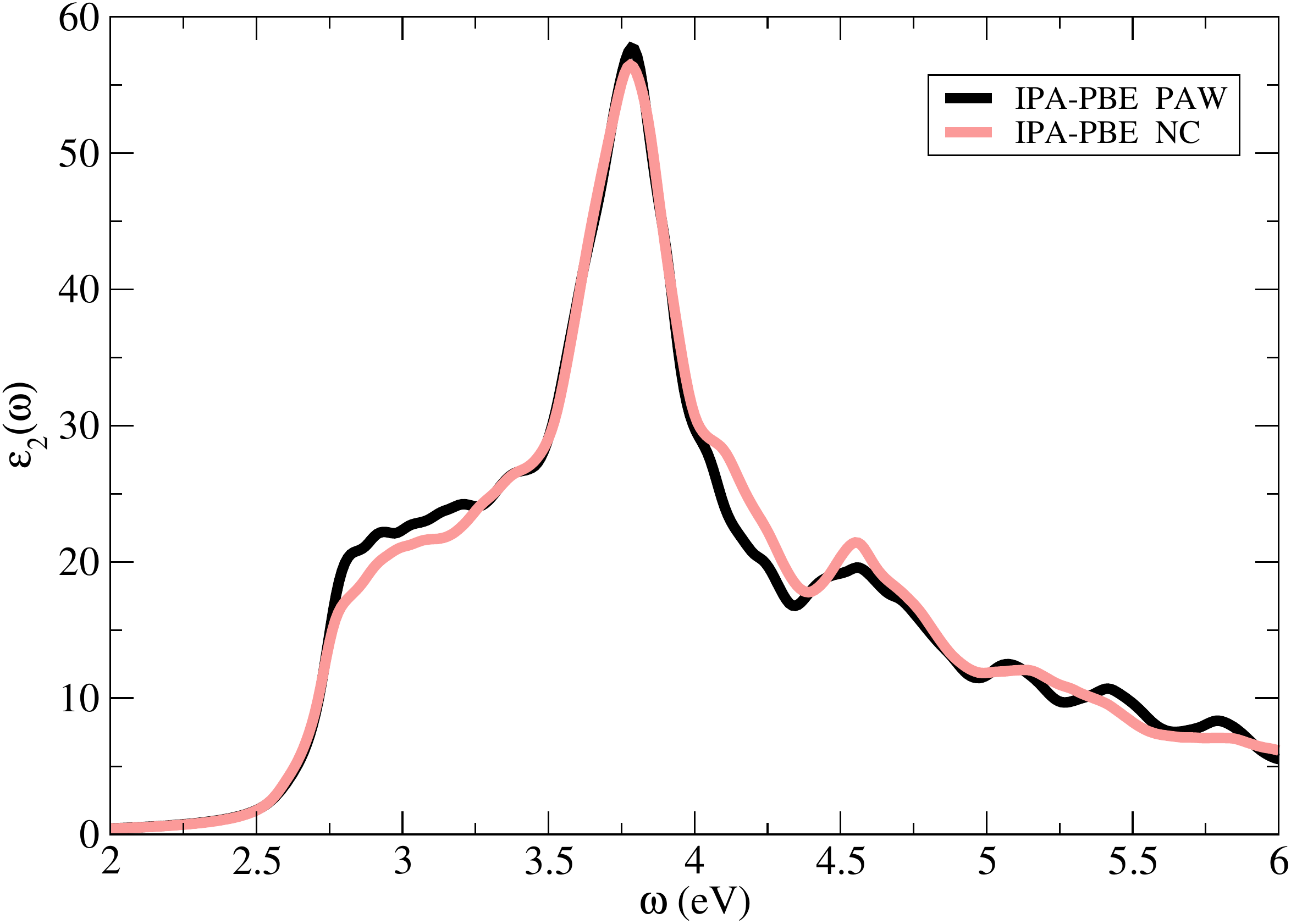}
\includegraphics[width=0.97\linewidth]{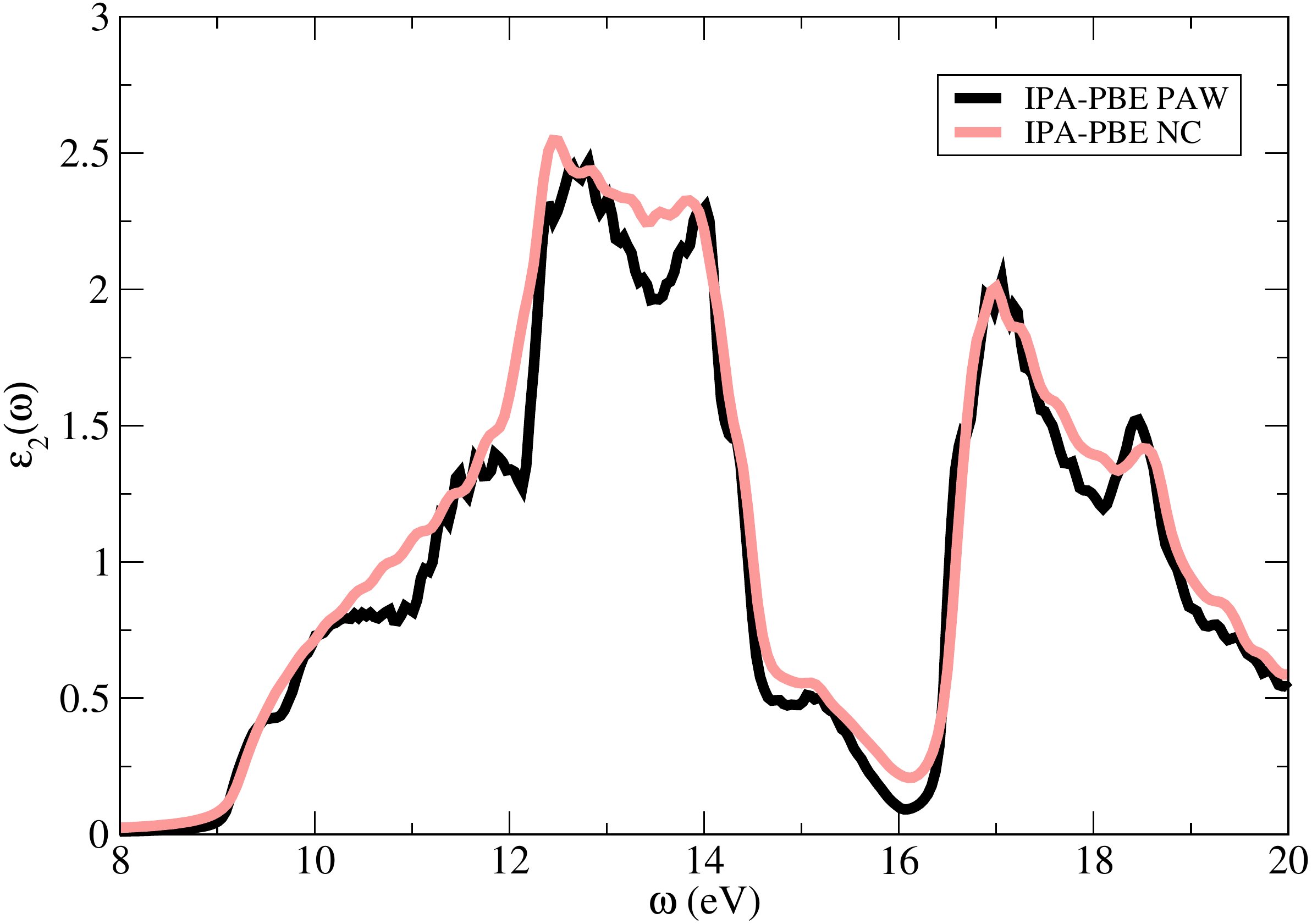}
\caption{$\varepsilon_2$ for Si (top panel) and LiF (bottom panel) calculated in IPA using NC and PAW pseudopotentials. }
\label{SiLiFIPAPBEVasp2light}
\end{figure}
\end{center}

In Fig.~(\ref{SiLiFIPAPBEVasp2light}) we show IPA-PBE for Si and LiF. We observe that the agreement is excellent between NC and PAW pseudopotentials. \cite{PhysRevB.63.125108} This implies that the differences we can observe in the spectra calculated with an higher level of theory are only due to the relevance of the TDDFT and TDGKSDFT kernels for the description of the excitons.
%====================================================================

%\begin{figure}[tbp]
%  \begin{subfigure}[b]{0.49\textwidth}
%    \includegraphics[width=\textwidth]{SiIPAPBEVasp2light-eps-converted-to.pdf}
%    \caption{Si}
%    \label{fig:}
%  \end{subfigure}
%  \hfill
%  \begin{subfigure}[b]{0.5\textwidth}
%    \includegraphics[width=\textwidth]{LiFIPAPBEVasp2light-eps-converted-to.pdf}
%    \caption{LiF}
%    \label{fig:f2}
%  \end{subfigure}
%  \caption{$\varepsilon_2$ for Si (top panel) and LiF (bottom panel) calculated in IPA using NC and PAW pseudopotentials.}
%  \label{SiLiFIPAPBEVasp2light}
%\end{figure}
%====================================================================

\begin{center}
\begin{figure}
\includegraphics[width=0.99\linewidth]{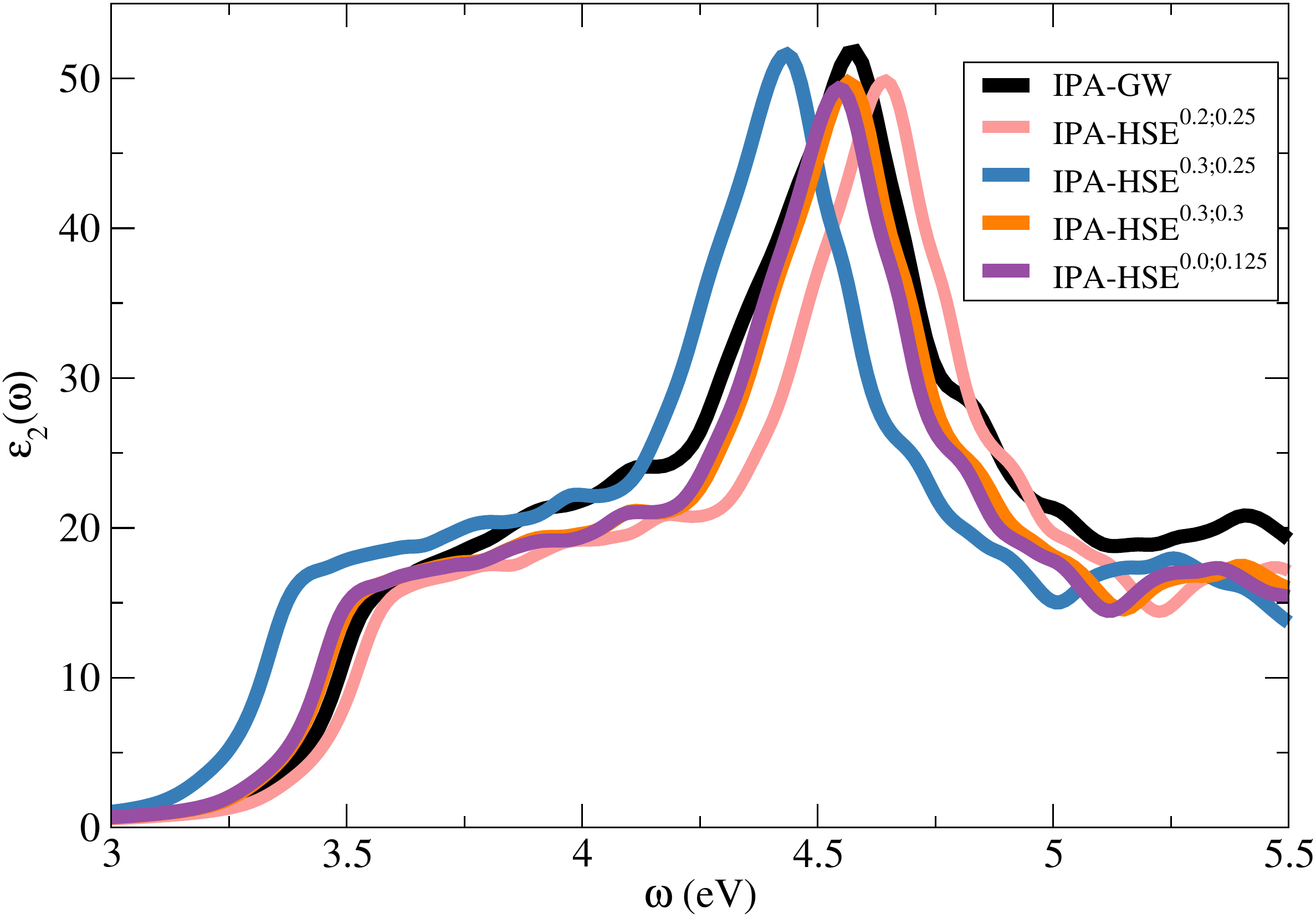}
\caption{$\varepsilon_2$ for Si calculated in IPA using GW, IPA-HSE$^{sr,0.2;0.25}$, HSE$^{sr,0.3;0.25}$, HSE$^{sr,0.3;0.3}$ and HSE$^{sr,0.0;0.125}$ and PAW pseudopotentials.}
\label{SiQuasiparticleFig}
\end{figure}
\end{center}

\begin{center}
\begin{figure}
\includegraphics[width=0.99\linewidth]{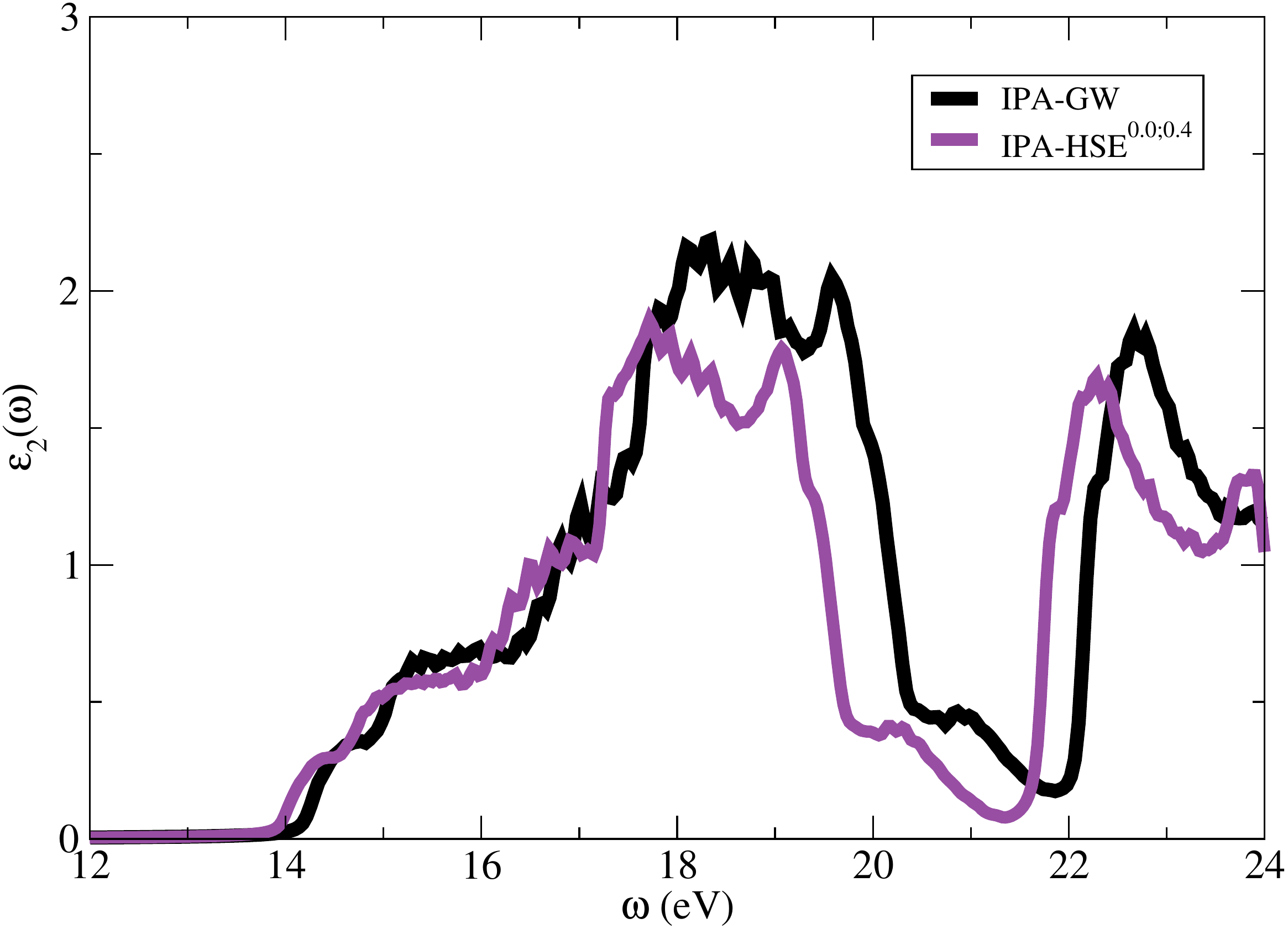}
\caption{$\varepsilon_2$ for LiF calculated in IPA using GW, IPA-HSE$^{sr,0.0;0.4}$ with PAW pseudopotentials.}
\label{LiFQuasiparticleFig}
\end{figure}
\end{center}

In Fig.~(\ref{SiQuasiparticleFig}) and in Fig.~(\ref{LiFQuasiparticleFig}) we compare IPA-GW and IPA-HSE which are IPA spectra calculated respecitvely on top of GW electronic structure and HSE$^{sr,\mu;\alpha}$ electronic structure where $\mu$ and $\alpha$ values are those that reproduce the GW gaps (see Table (\ref{Sigaps}) and Table (\ref{LiFgaps})).

For Si the calculations are consistent as shown in Fig.~(\ref{SiQuasiparticleFig}). The trend is the same we observed for the electronic gaps of Table (\ref{Sigaps}). In fact, IPA-HSE$^{sr,0.3;0.25}$ is slightly lower than IPA-HSE$^{sr,0.2;0.25}$ due to a larger value of $\mu$. Instead, using the same value of $\mu=0.3$ but an higher value of $\alpha=0.3$ as in IPA-HSE$^{sr,0.3;0.3}$, the spectrum shifts at higher energy due to a larger percentage of nonlocal HF exchange. Also in the case of LiF we found a good agreement as shown in Fig.~(\ref{LiFQuasiparticleFig}), where we compared the IPA-GW spectrum with IPA-HSE$^{sr,0.0;0.4}$.
%==============================================================
%===============================================================
%===============================================================
The spectra of Fig.~(\ref{SiQuasiparticleFig}) and Fig.~(\ref{LiFQuasiparticleFig}) do not include excitonic effects as IPA is the lowest level of approximation for the calculation of optical spectra.  

\begin{center}
\begin{figure}
\includegraphics[width=0.99\textwidth]{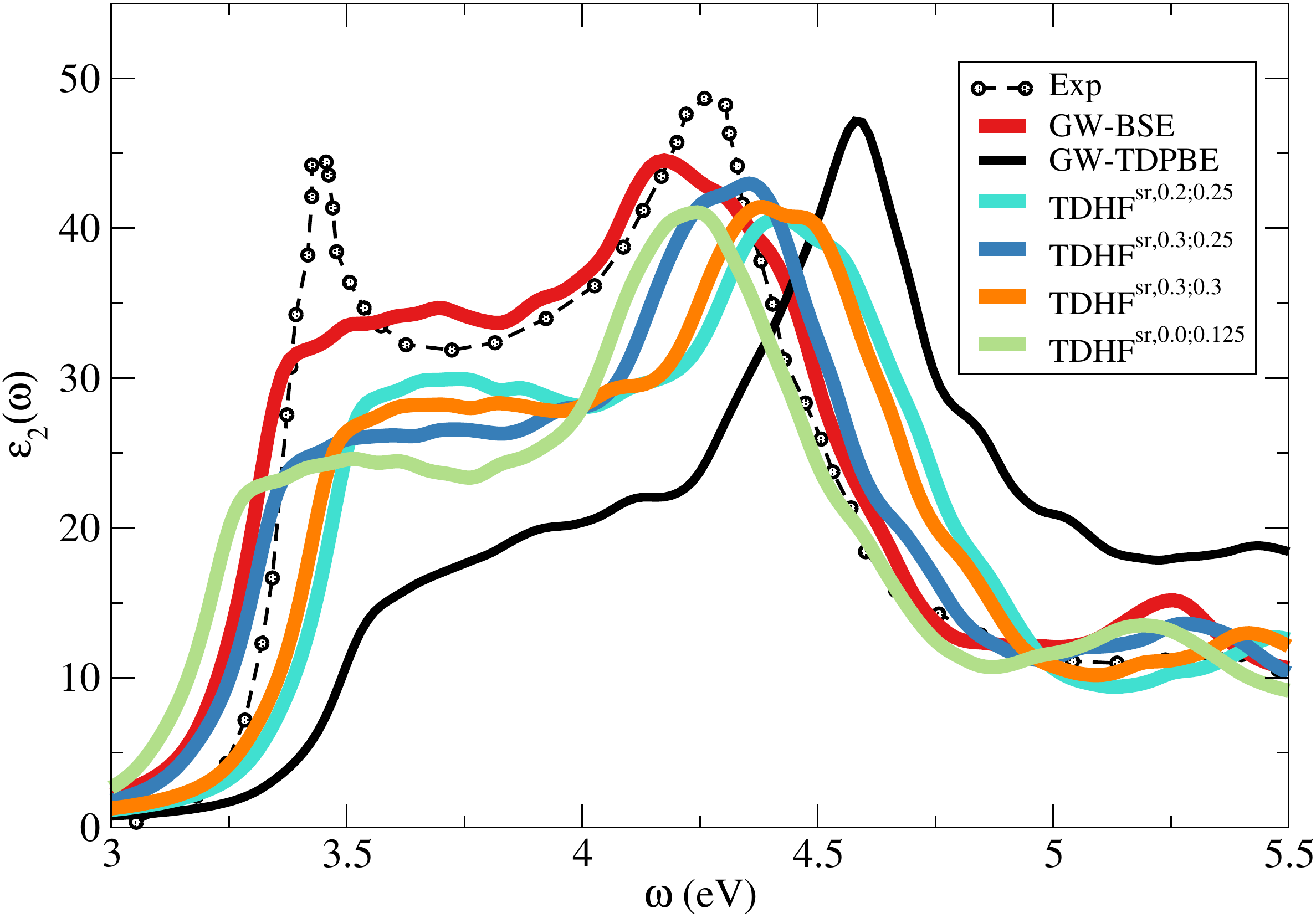}
\caption{Comparison of experimental $\varepsilon_2$ for Si with GW-BSE, GW-TDPBE, TDHF$^{sr,0.2;0.25}$, TDHF$^{sr,0.3;0.25}$, TDHF$^{sr,0.3;0.3}$ and TDHF$^{sr,0.0;0.125}$ and PAW pseudopotential. Experiment is from Ref. \cite{PhysRevB.36.4821}.}
\label{SiGWRPABSETDHF}
\end{figure}
\end{center}

As already pointed out in \cite{rein+02prl}, we show in Fig.~(\ref{SiGWRPABSETDHF}) the excellent agreement of GW-BSE with the experimental spectrum of Si. GW-TDPBE, as expected, is not able to reproduce excitonic effects and it only slightly improves the spectrum  with respect to GW-RPA (see Refs. \cite{bott+04prb,PhysRevLett.91.056402,PhysRevB.87.205143}). In fact, in TDPBE the exchange-correlation kernel is PBE which has not the proper spatial nonlocality. \cite{onid+02rmp} TDHF$^{\text{sr},\mu;\alpha}$ optical spectra have a reasonable shape but the intensity of the first peak around 3.5 eV is  too low. In order to increase the intensity of this peak, we need to increase the percentage of the nonlocal HF exchange, as can be seen by comparing TDHF$^{sr,0.3;0.25}$ with TDHF$^{sr,0.2;0.25}$. Otherwise, another strategy would be to increase the value of the mixing parameter $\alpha$ as observed by comparing TDHF$^{sr,0.3;0.25}$ with TDHF$^{sr,0.3;0.3}$. However, we believe that the use of short-range HF exchange has not the necessary flexibility to improve further the spectrum. In fact, increasing $\alpha$ or $\mu$ would change also the energy position of the peaks. 

TDHF$^{\text{sr},0.0;0.125}$ contains the full range nonlocal HF exchange. The $\alpha=0.125$ we have chosen, permits to be consistent with the previous step, i.e. a correct electronic structure. However, this value of $\alpha$ is still too small to correctly reproduce the experimental spectrum. Similar results were also obtained by Yang {\it et al.} \cite{PhysRevB.92.035202} using for $\alpha$ the value of the inverse RPA dielectric constant ($\sim$0.08). A better description of the first peak could be done by increasing the $\alpha$ value, but also in this case this would cause a change in the position of the energy peaks. 

\begin{center}
\begin{figure}
\includegraphics[width=0.99\textwidth]{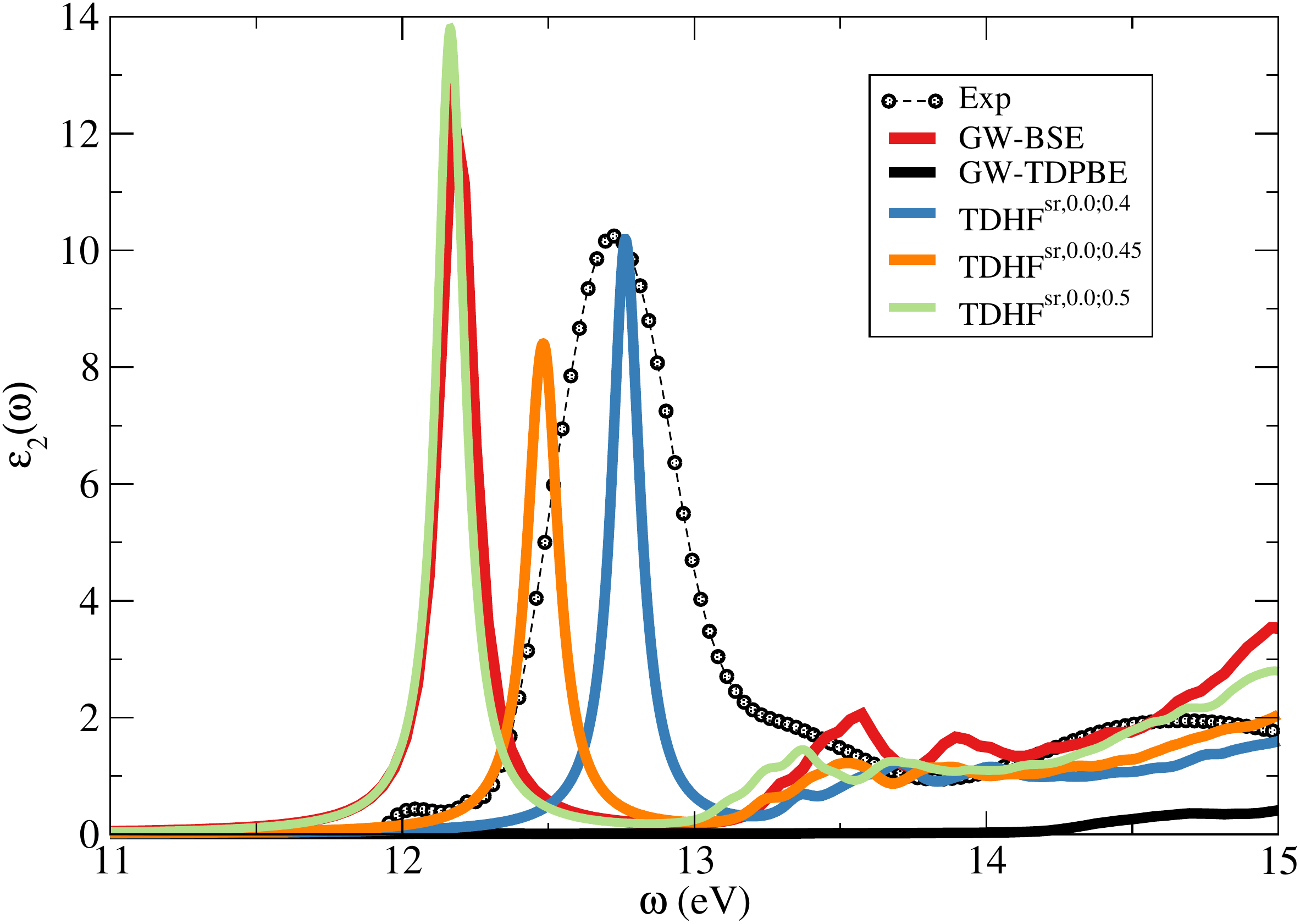}
\caption{Comparison of experimental $\varepsilon_2$ for LiF with GW-BSE, GW-TDPBE, TDHF$^{sr,0.0;0.4}$, TDHF$^{sr,0.0;0.45}$ and TDHF$^{sr,0.0;0.5}$ and PAW pseudopotentials. Experiment is from Ref. \cite{Roessler:67}.}
\label{LiFSiGWRPABSETDHF}
\end{figure}
\end{center}

In the case of LiF, the GW-BSE reproduces an excitonic peak of 12.2 eV, which is slightly lower \cite{PhysRevResearch.2.013091} than the experimental peak of 12.75 eV, as shown in Fig.~(\ref{LiFSiGWRPABSETDHF}). Instead, as expected, GW-TDPBE can not reproduce the excitonic peak. TDHF$^{sr,0.0;0.4}$ gives an excellent agreement with the energy position of the experimental exciton. By increasing the value of $\alpha$ we include more nonlocal HF exchange and therefore the exciton is more strongly bound as we have shown for TDHF$^{sr,0.0;0.45}$ and TDHF$^{sr,0.0;0.5}$, see Fig.~(\ref{LiFSiGWRPABSETDHF}).
 
\begin{center}
\begin{figure}
\includegraphics[width=0.99\textwidth]{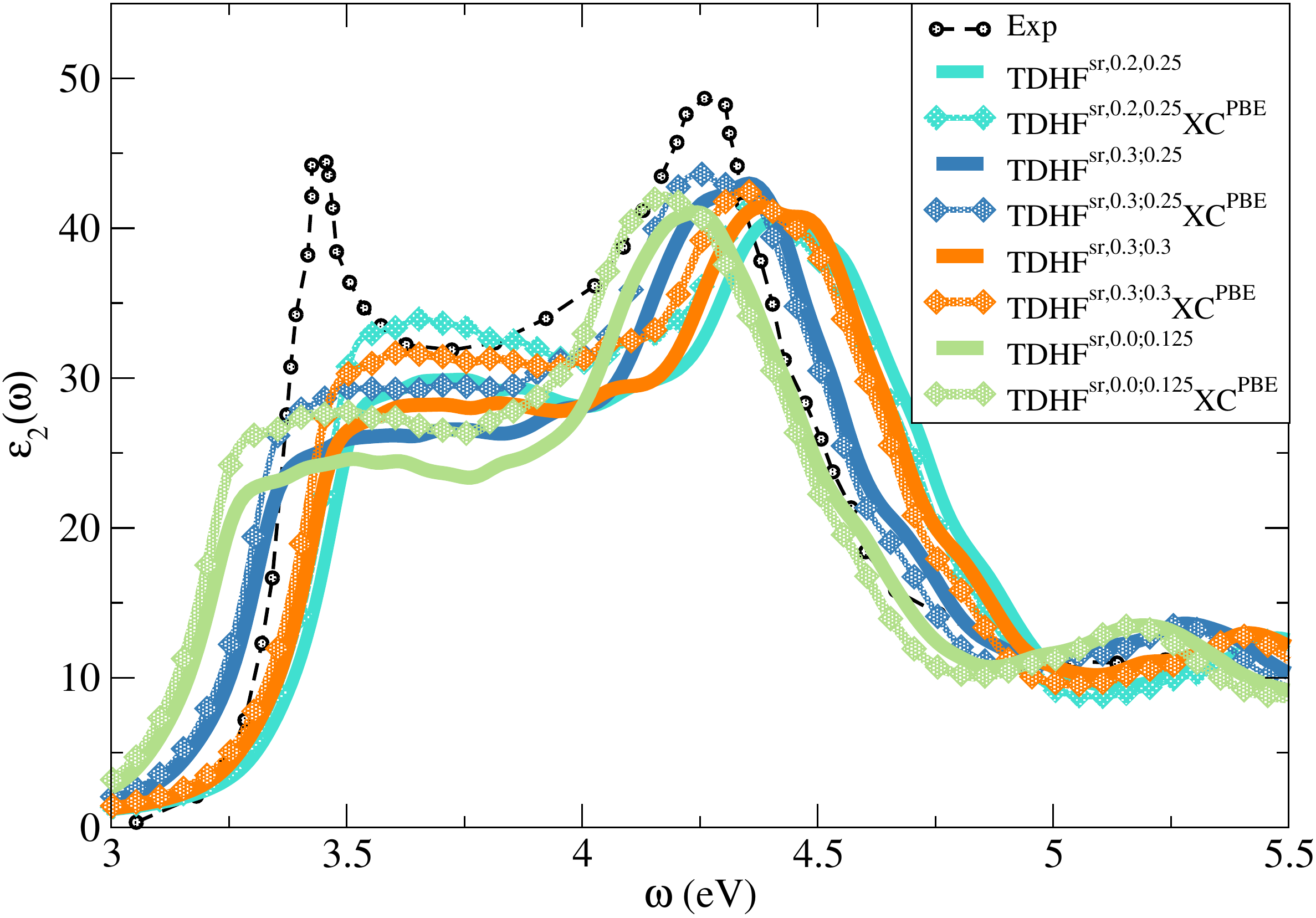}
\caption{$\varepsilon_2$ for Si : effect of inclusion of a fraction of PBE exchange-correlation. PAW pseudopotential has been used. Experiment is from Ref. \cite{PhysRevB.36.4821}.}
\label{SiTDPBETDHFFXCPBE}
\end{figure}
\end{center}

\begin{center}
\begin{figure}
\includegraphics[width=0.99\textwidth]{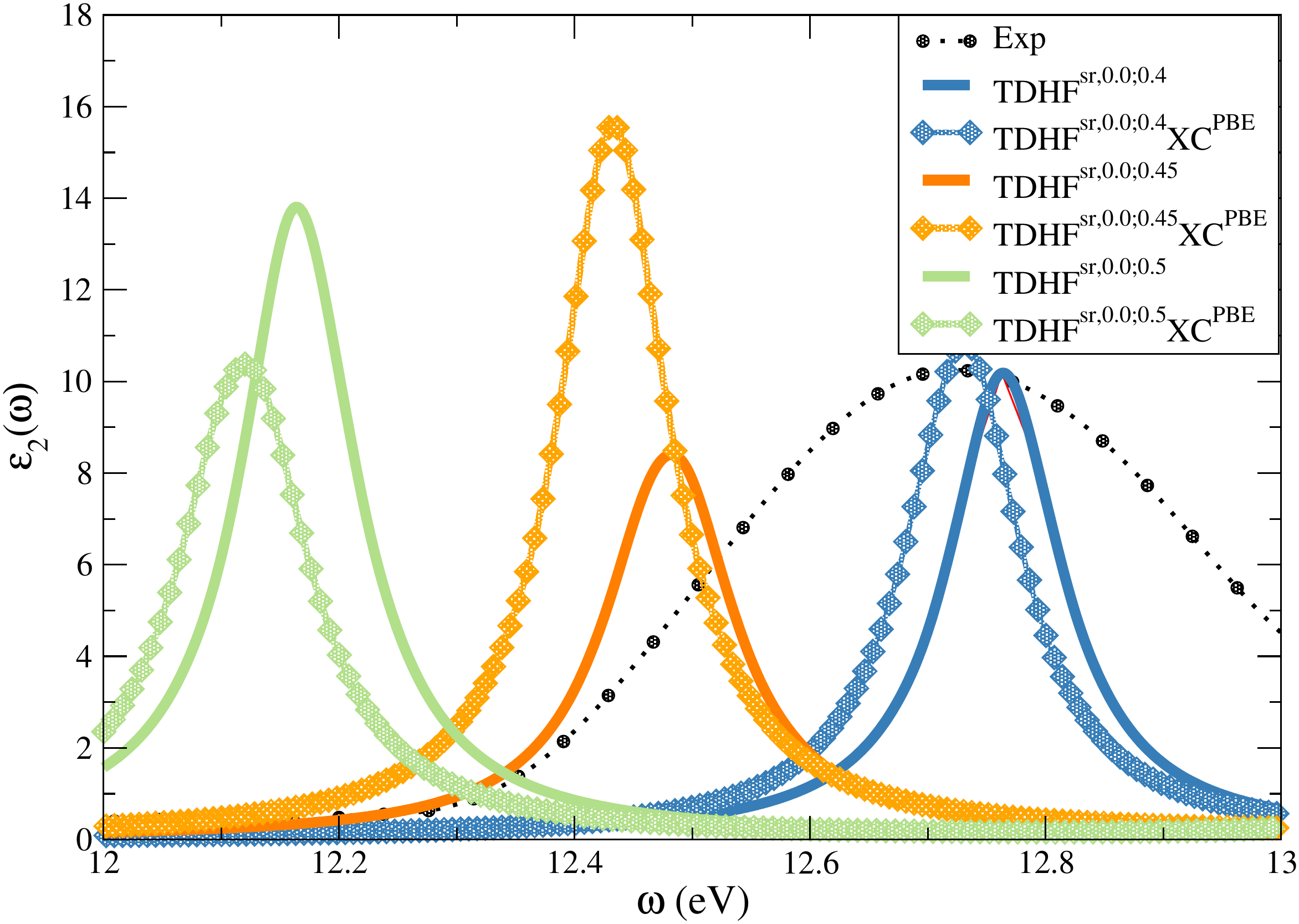}
\caption{$\varepsilon_2$ for LiF : effect of inclusion of a fraction of PBE exchange-correlation. PAW pseudopotentials have been used. Experiment is from Ref. \cite{Roessler:67}.}
\label{LiFTDPBETDHFFXCPBE}
\end{figure}
\end{center}

The comparison between TDHF$^{sr,\mu;\alpha}$ and TDHF$^{sr,\mu;\alpha}$XC$^{\text{PBE}}$ is in Fig.~(\ref{SiTDPBETDHFFXCPBE}) for Si and in Fig.~(\ref{LiFTDPBETDHFFXCPBE}) for LiF. From Eq.~(\ref{tdhf}) and Eq.~(\ref{tdhfpbe}) the difference between these kernels is the addition to the nonlocal HF exchange of a fraction of the semilocal exchange PBE $(1-\alpha)f^{\text{x,PBE}}$ and the PBE correlation $f^{\text{c,PBE}}$. In the case of Si, adding a fraction of semilocal exchange increases the intensity of the first peak around 3.5 eV, therefore, improving the agreement with the experiment. The energy position of the peak is not changed. Instead, in the case of LiF the energy position of the peak is slightly shifted to lower energy and the intensity of the peak changes. However, concerning the peak intensity we did not find a clear trend.

\begin{center}
%=====================================================================
\begin{figure*}
\includegraphics[width=0.99\textwidth]{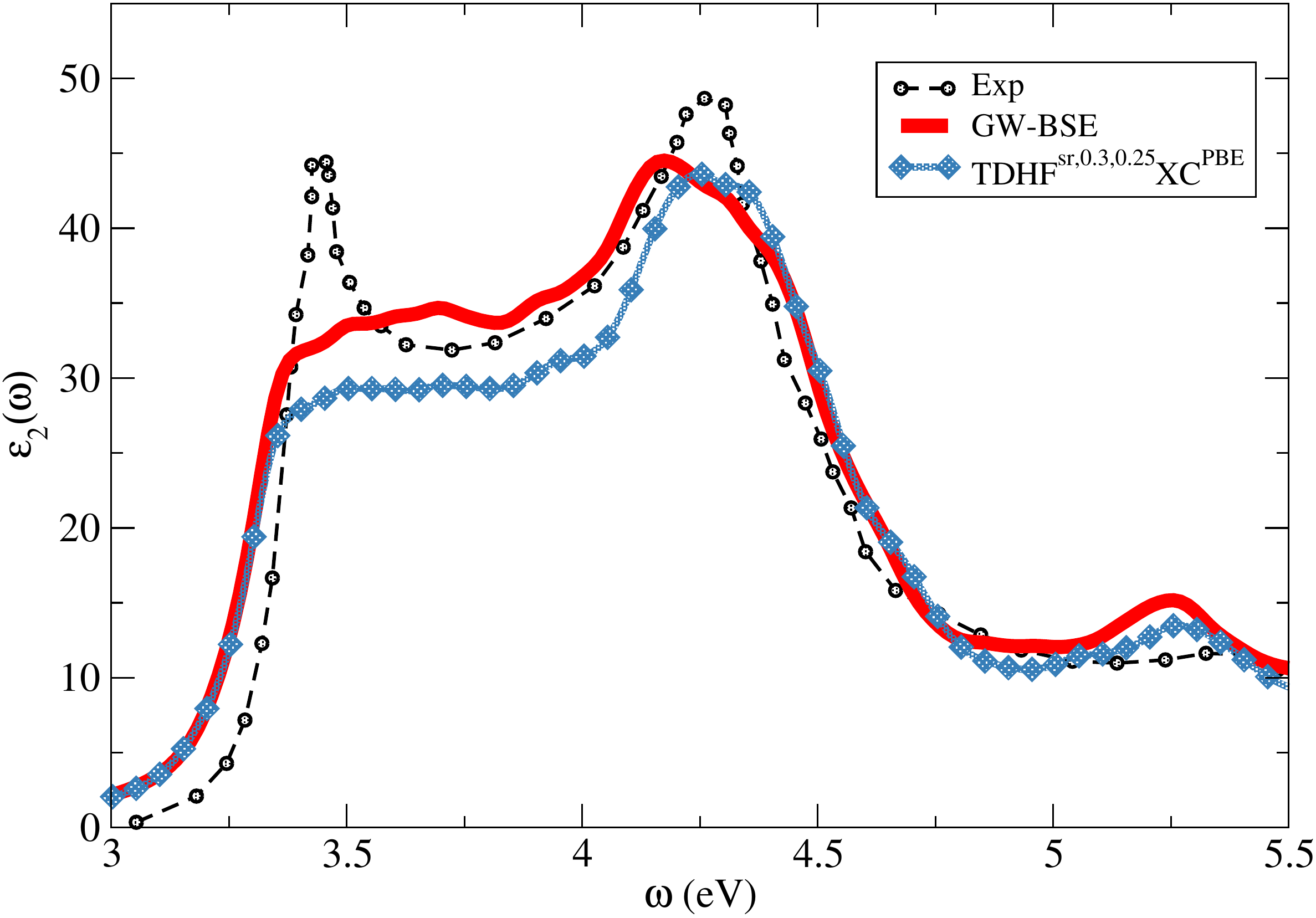}
\caption{Comparison of experimental $\varepsilon_2$ for Si with GW-BSE and TDHF$^{sr,0.3;0.25}$XC$^{\text{PBE}}$. PAW pseudopotential has been used. Experiment is from Ref. \cite{PhysRevB.36.4821}.}
\label{SiGWBSETDHFFXCPBE}
\end{figure*}
\end{center}

\begin{center}
\begin{figure*}
\includegraphics[width=0.99\textwidth]{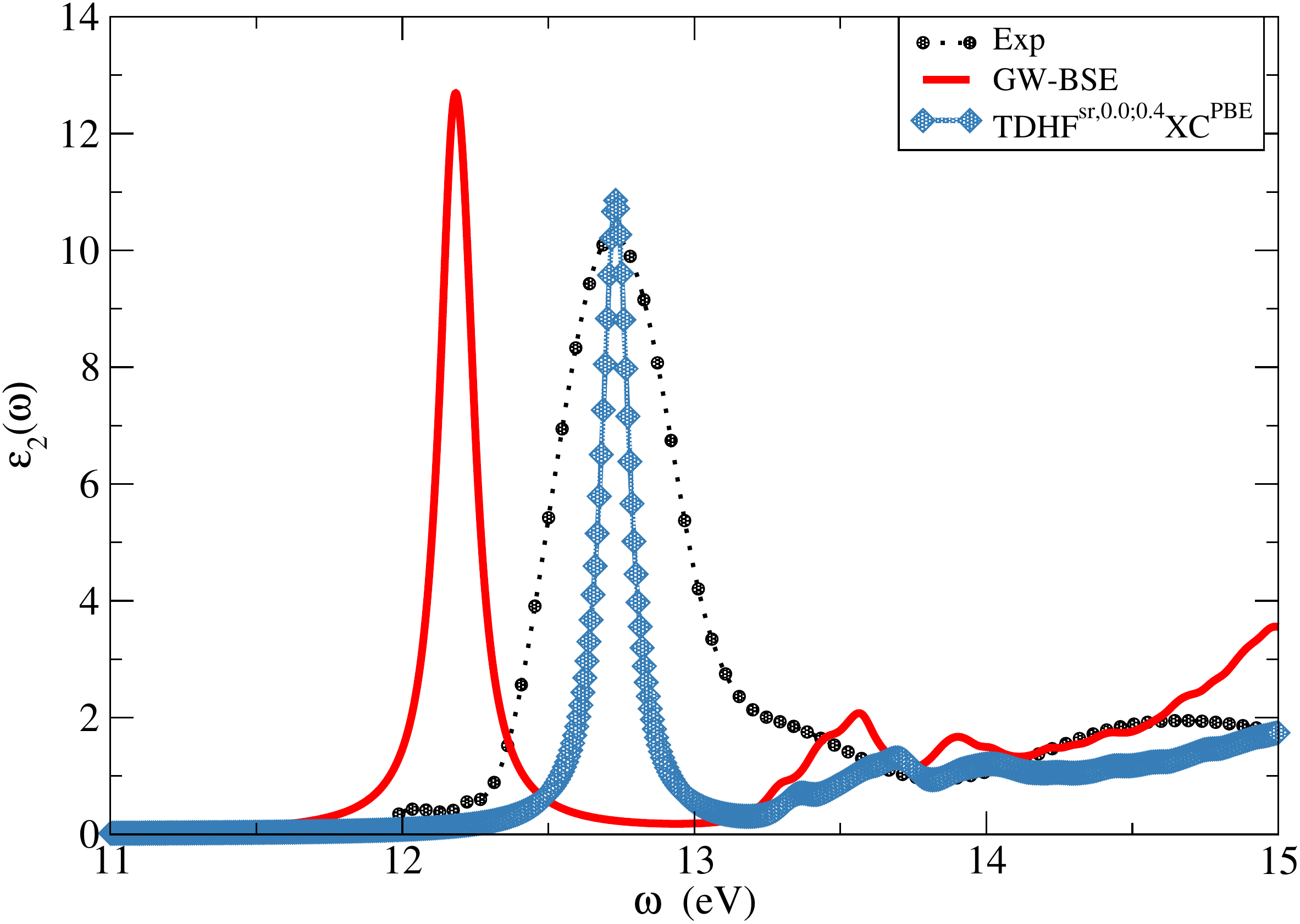}
\caption{Comparison of experimental $\varepsilon_2$ for LiF with GW-BSE and TDHF$^{sr,0.0;0.4}$XC$^{\text{PBE}}$. PAW pseudopotentials have been used. Experiment is from Ref. \cite{Roessler:67}.}
\label{LiFGWBSETDHFFXCPBE}
\end{figure*}
\end{center}

In Fig.~(\ref{SiGWBSETDHFFXCPBE}) and Fig.~(\ref{LiFGWBSETDHFFXCPBE}) we finally present the TDHF$^{sr,\mu;\alpha}$XC$^{\text{PBE}}$ spectra which give the best agreement with experiment and we compare them to the GW-BSE spectra.

%=====================================================================
%=====================================================================
%=====================================================================
%=====================================================================
%=====================================================================
%=====================================================================
\begin{center}
\begin{figure*}
\includegraphics[width=0.99\textwidth]{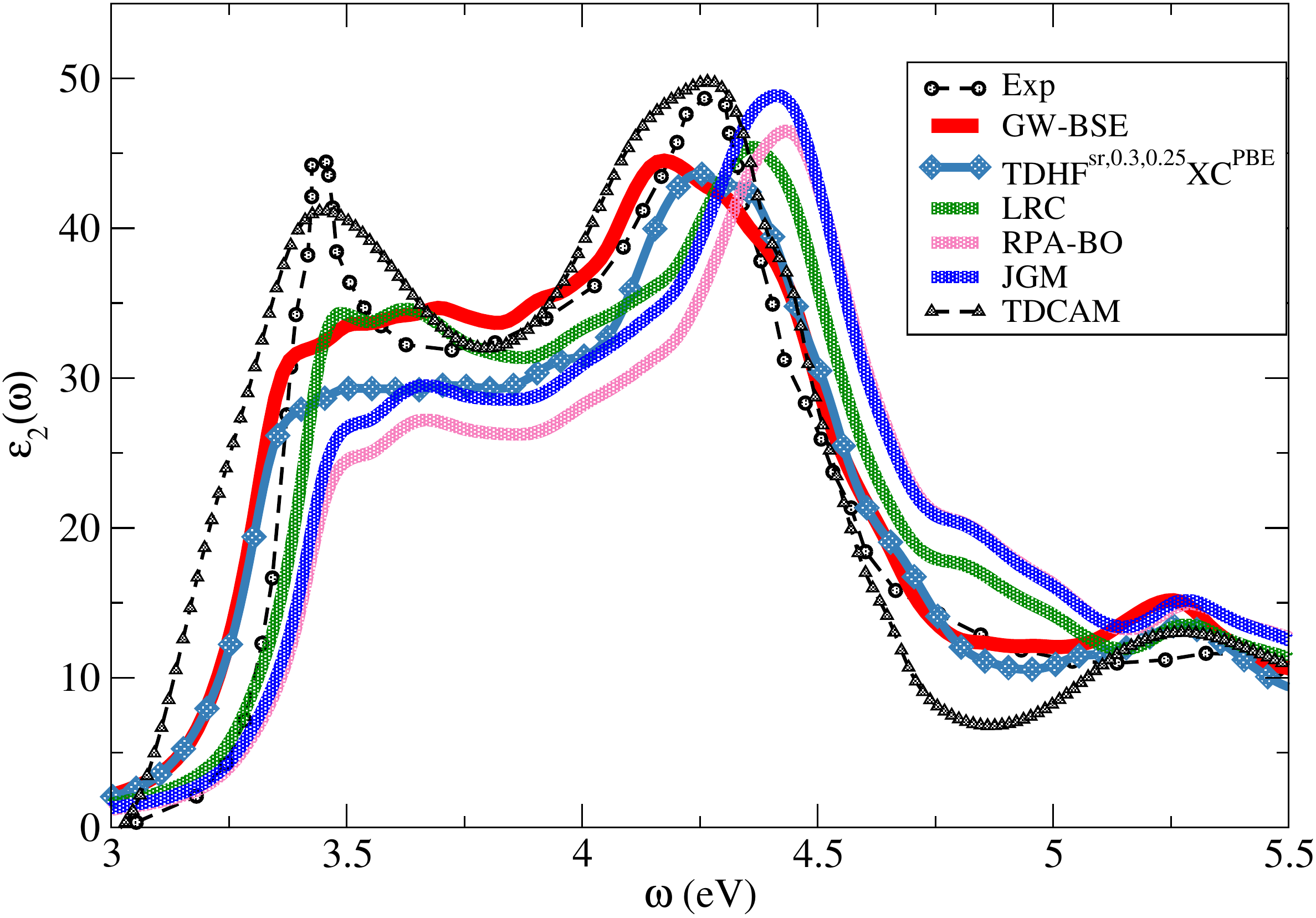}
\caption{Comparison of experimental $\varepsilon_2$ for Si with GW-BSE, TDHF$^{sr,0.3;0.3}$XC$^{\text{PBE}}$  (PAW pseudopotential) and with the long-range corrected kernels : LRC, RPA-BO and JGM (NC pseudopotential). TDCAM is from Ref. \cite{PhysRevB.92.081204}. Experiment is from Ref. \cite{PhysRevB.36.4821}.}
\label{SiGWBSETDHFFXCPBELRC}
\end{figure*}
\end{center}
%=====================================================================
\begin{center}
\begin{figure*}
\includegraphics[width=0.99\textwidth]{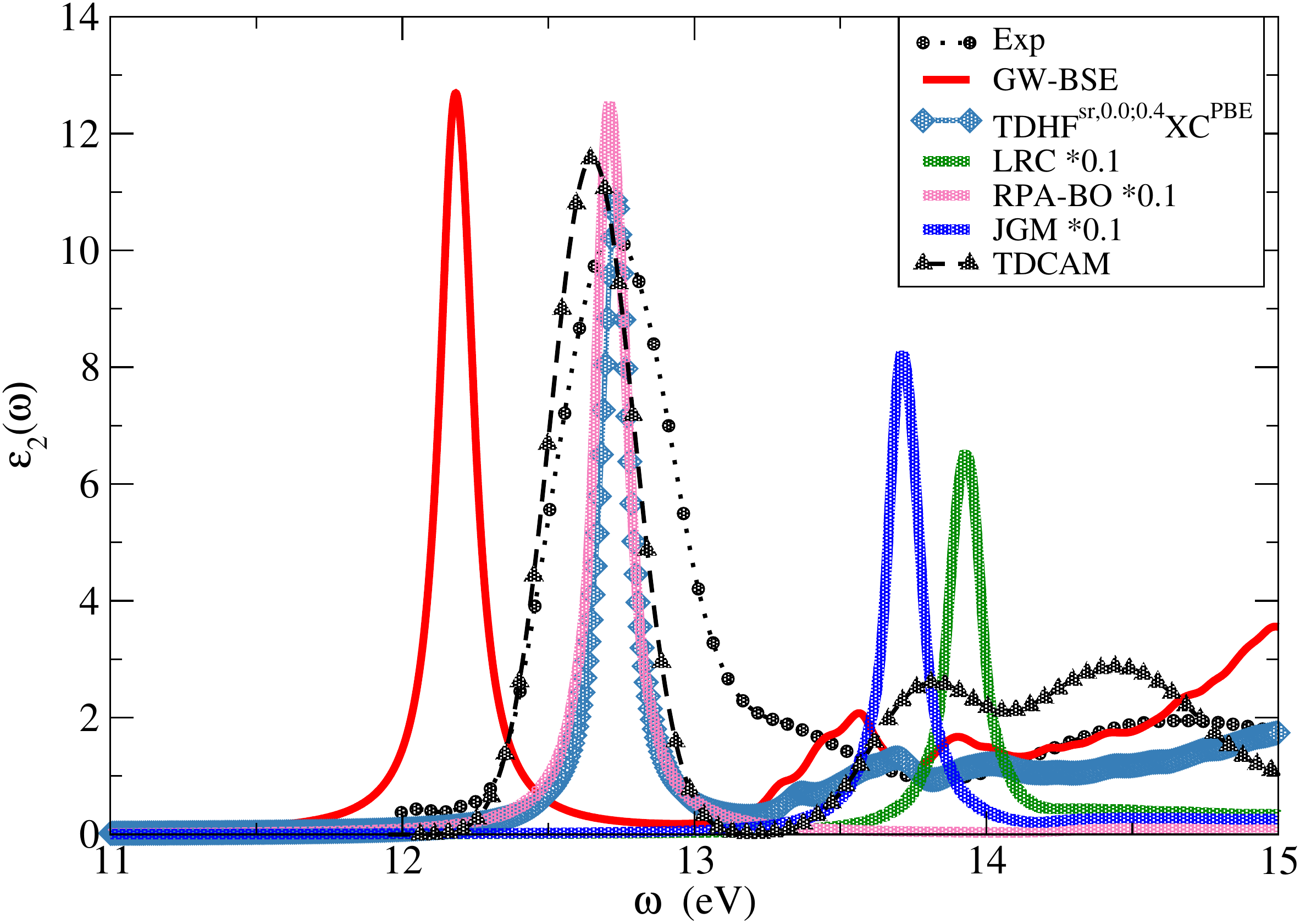}
\caption{Comparison of experimental $\varepsilon_2$ for LiF with GW-BSE, TDHF$^{sr,0.3;0.3}$XC$^{\text{PBE}}$  (PAW pseudopotentials) and with the long-range corrected kernels (NC pseudopotentials): LRC, RPA-BO and JGM. TDCAM is from Ref. \cite{PhysRevB.92.081204}. Experiment is from Ref.\cite{Roessler:67}.}
\label{LiFGWBSETDHFFXCPBELRC}
\end{figure*}
\end{center} 
%=====================================================================

Finally, for Si, we compare in Fig.~(\ref{SiGWBSETDHFFXCPBELRC}) the selected TDHF$^{sr,0.3;0.25}$XC$^{\text{PBE}}$ spectrum to TDDFT with long-range corrected kernels with a scissor shift of 0.7 eV (see Table~(\ref{Sigaps})).

We show the results for LRC ($\alpha^{\text{LRC}}=0.20$), RPA-BO ($\alpha^{\text{RPA-BO}}=0.13$) and JGM ($\alpha^{\text{JGM}}=0.12$) kernels. \cite{gaurJCP2019} For the long-range corrected kernels a higher value of the $\alpha$ parameter can be interpreted as if a higher nonlocal HF long-range contribution was included. This contribution is higher in LRC than in RPA-BO and JGM kernels. This is the reason for which LRC better describes the spectrum around 3.5 eV.

However, despite the better behaviour of LRC kernel, we want to point out that RPA-BO and JGM kernels do not require any adjustable parameter, which is an enormous advantage as they can be applied to any kind of materials.

The TDHF$^{sr,0.3;0.25}$XC$^{\text{PBE}}$ gives a reasonable Si spectrum, similar to the spectra from the RPA-BO and the JGM kernels.

We add to the comparison the result from the range-separated CAM proposed by Rafaely-Abramson {\it et al.} \cite{PhysRevB.92.081204} which contains a fraction of long-range nonlocal HF exchange. The approach of Rafaely-Abramson {\it et al.} \cite{PhysRevB.92.081204} is in excellent agreement with experiment and seems also to improve with respect to GW-BSE. However, this approach contains 3 parameters $\alpha$, $\beta$ and $\mu$. The parameter $\alpha$ is the amount of short-range exact exchange, $\beta$ is calculated as $\alpha+\beta=1/\varepsilon_0$ and $\mu$ is the range-separation parameter. To obtain this excellent agreement, $\mu$ was optimally tuned in order to reproduce the electronic gap, and $\alpha$ and $\beta$ are obtained from the material's dielectric constant $\varepsilon_0$. They used $\mu=0.11$ Bohr$^{-1}$, $\alpha=0.2$ and $\varepsilon_0=12$.

Comparing the method with a fraction of short-range nonlocal HF exchange ($\mu=0.3$ Bohr$^{-1}$ and $\alpha=0.25$) to the one with long-range nonlocal HF exchange ($\mu=0.11$ Bohr$^{-1}$, $\alpha=0.2$), and considering that the mixing parameter $\alpha$ is of the same order of magnitude, we observe that the range-separation parameter $\mu$ is larger when a fraction of short-range is used. This is reasonable as the role of $\mu$ is opposite between short and long-range.  

However, using TDHF$^{sr,0.3;0.25}$XC$^{\text{PBE}}$ (short-range), it is not possible to obtain the same agreement with experiment that is reproduced when a fraction of long-range nonlocal HF exchange is included. To obtain the same performance of the long-range scheme, the value of $\mu$ should be increased. However, this would cause a shift of the excitonic peaks to lower energy and the spectrum will be wrong.

In Fig.~(\ref{LiFGWBSETDHFFXCPBELRC}), for LiF, we compare TDHF$^{sr,0.0;0.4}$XC$^{\text{PBE}}$ with TDDFT with long-range corrected kernels with a scissor shift of 5.0 eV (see Table~(\ref{LiFgaps})).

We used $\mu=0.0$ Bohr$^{-1}$ and $\alpha=0.4$ as we need the full range nonlocal HF exchange to reproduce the experimental spectrum. In fact, we did not find any finite values of $\mu$ different from zero for which using only a fraction of short-range nonlocal HF exchange it would be possible to reproduce the experimental spectrum.

We show LRC ($\alpha^{\text{LRC}}=8.0$), RPA-BO ($\alpha^{\text{RPA-BO}}=8.8$) and JGM ($\alpha^{\text{JGM}}=7.93$) kernels. \cite{gaurJCP2019} The TDHF$^{sr,0.0;0.4}$XC$^{\text{PBE}}$ and RPA-BO are in excellent agreement with the energy position of the excitonic peak. However, RPA-BO, as well as LRC and JGM overestimate the peak intensity, which in Fig.~(\ref{LiFGWBSETDHFFXCPBELRC}) has been multiplied by 0.1 in order to compare the theoretical approaches. Furthermore, we observe that the energy of the JGM peak is around 1 eV higher than the result presented in the original work of Trevisanutto {\it et al.} \cite{PhysRevB.87.205143}. This is due to the different scissor value taken to correct the energies.

We add to this comparison also the result from the range-separated CAM proposed by Rafaely-Abramson {\it et al.} \cite{PhysRevB.92.081204} which is in excellent agreement with the experiment and also improve with respect to the GW-BSE. They used $\mu=0.58$ Bohr$^{-1}$, $\alpha=0.2$ and $\epsilon_0=1.9$.

\section{Conclusion}
\label{conclusion}

We compared the performance of TDGKSDT range-separated hybrid functionals and TDDFT long-range corrected kernels for the description of excitons in solids. The comparison was illustrated for the case of Si and LiF, representative of continuum and strong excitons.

We studied hybrid functionals with a fraction of short-range nonlocal HF exchange. For Si, by optimally tuning $\mu$ and $\alpha$, it is possible to reproduce the satisfactory experimental spectrum. Instead, for LiF we did not find any finite values of $\mu$, different from zero, for which is possible to reproduce the experimental spectrum. In the case of LiF we need to use the (full range) nonlocal HF exchange ($\mu=0.0$) in order to satisfactory reproduce the experiment. Therefore, exchange is much more important for strong excitons than for weak ones.

We also studied the long-range corrected kernels: LRC \cite{rein+02prl}, RPA-BO \cite{PhysRevLett.114.146402} and JGM \cite{PhysRevB.87.205143}. These kernels perform comparably to hybrid functionals with short-range nonlocal HF exchange. Except that for LiF the intensity of the excitonic peak is strongly overestimated.

We included in our discussion the hybrid scheme of Refs.~\cite{PhysRevB.92.081204,PhysRevMaterials.3.064603} which has a long-range nonlocal HF exchange component. This approach has an excellent agreement with experiment for both Si and LiF and it also seems to improve with respect to GW-BSE. This approach is the most flexible.

From this comparison it appears that the hybrid scheme with long-range nonlocal HF exchange performs better than the hybrid scheme with short-range nonlocal HF exchange. We believe that for Si, and therefore for weak excitons, it is the lack of long-range component of the nonlocal HF exchange which causes a not yet excellent description of the exciton around 3.4 eV. The situation is even worse for LiF, and therefore for strong excitons, where the short-range separation demonstrated not to work.

The main difficulty of using range-separated hybrid functionals is their dependence on parameters that have to be chosen and which strongly depends on the material. A general strategy to find these parameters is needed.

Moreover, despite the promising behaviour of range-separated schemes, long-range kernels continue to be attractive. In fact, the computational cost is lower and a kernel such as RPA-BO does not require any adjustable parameters.

\begin{acknowledgments}

This work was performed using HPC resources from GENCI-IDRIS  Grant 2021-x2021082131 and Grant 2021-x202109544.

\end{acknowledgments}

%\appendix

%%%\bibliography{bib}

\end{document}